\documentclass[preprint,showpacs,preprintnumbers,amsmath,amssymb]{revtex4}
\usepackage{graphicx}
\usepackage{dcolumn}
\usepackage{bm}
\begin{document}
\title {The Spin Mass of an Electron Liquid}
\author {Zhixin Qian}
\altaffiliation[Also at: ]{Department of Physics  and State 
Key Laboratory for Mesoscopic Physics, Peking University, Beijing 100871, China.}
\author{Giovanni Vignale}
\affiliation{ Department of Physics and Astronomy, 
University of Missouri, Columbia, Missouri 65211, USA} 
\author{D. C. Marinescu}
\affiliation{Department of Physics, Clemson University, Clemson, South Carolina 29634, USA}
\date{\today}
\begin{abstract}
We show that in order to calculate correctly the {\it spin current} 
carried by a quasiparticle in an electron liquid  one must use an
effective ``spin mass" $m_s$,  that is larger  than both the band 
mass, $m_b$, which determines the charge current, and 
the quasiparticle effective mass $m^*$, which determines the heat capacity. 
We present  microscopic calculations of  $m_s$ in 
a paramagnetic electron liquid in three and two dimensions,  
showing that the mass enhancement $m_s/m_b$ can be a very significant effect. 
\end{abstract}
\maketitle

The recent explosion of interest in metal and semiconductor 
spintronics~\cite{Wolf,Awschalom} has brought into  sharp focus 
the basic problem of calculating the {\it spin current} carried 
by a nonequilibrium electronic system. 
The standard approach is to solve the Boltzmann equation 
for the  nonequilibrium distribution function;  but this 
is not sufficient when many-body effects due to electron-electron 
interactions need to be taken into account.   In fact, electronic correlations 
are particularly strong in low-dimensional systems,   
such as magnetic semiconductor films and wires, which are currently 
being considered for the realization of spin transistors~\cite{Ruster03}.  
One might hope to take care of
the  many-body effects  by solving, instead  of the Boltzmann equation, 
the Landau-Silin transport equation for quasiparticles~\cite{NP}.  
But even this is not sufficient, since the transport equation {\it per se} 
does not tell us how to connect the quasiparticle distribution 
function to the spin-current.   The key question, which seems 
to have been  overlooked so far in the growing 
literature on spin transport~\cite{zhangrashba}, 
is also a very basic one, namely,  what is the spin-current carried by a single 
quasiparticle of momentum $\vec p$ and spin $\sigma$?  Without knowing 
the answer to this question it is not possible to calculate the spin 
current from  first-principles.  
In this paper we show that, in order to calculate the spin current correctly, 
one must recognize that the effective spin mass $m_s$, which determines  
the relation between the spin current and the quasiparticle momentum, 
is neither the band mass $m_b$ (which controlls the charge current), 
nor the quasiparticle mass $m^*$ (which controls the heat capacity),  
but rather a new many-body quantity, controlled by spin correlations.  
Our calculations show that the spin mass, in spite of uncertainties due to the approximate character of the many-body theory,  can be considerably larger than the bare band mass in a two-dimensional electron gas (by contrast, the quasiparticle effective mass is typically very close to the band mass).  Hence, the spin mass  will have to 
be taken into account whenever a quantitative comparison between theory 
and experiment is desired.

Let us begin by describing the physical origin of the spin mass.
The spin current, $\vec  j_s = \vec j_\uparrow - \vec j_\downarrow$,  
is defined as the difference  
of the up-spin and down-spin currents, 
$\vec j_\uparrow$ and $\vec j_\downarrow$,  
which in turn are defined as the expectation values of the operators
\begin{equation} \label{currentoperators}
\hat{\vec j}_ \sigma = \sum_{i=1}^N
\frac{\hat{\vec p}_i}{m_b} \frac{ 1 + \sigma \hat \sigma_{z,i}}{2}~
\end{equation}
in the appropriate nonequilibrium state.
Here $\sigma = 1$ for $\uparrow$ spins 
and  $\sigma = -1$ for $\downarrow$ spins, $m_b$ is the bare band mass,
$\hat{\vec p}_i$ is the canonical momentum operator
of the $i$-th electron,  $\hat \sigma_{z,i}$ is the 
Pauli matrix of the $z$-component of the spin  of 
the $i$-th electron, $\frac{1 + \sigma \hat \sigma_{i,z}}{2}$ is 
the projector on the $\sigma$-spin component  of 
the $i$-th electron,  and $N$ is the number of electrons.
Let us consider a many-body state, denoted by $|\vec p\sigma\rangle$,  
which contains a single quasiparticle of 
momentum $\vec p$ and spin $\sigma$. This state carries a total 
current $\vec j = \frac{\vec p}{m_b}$, whether or not interactions
are taken into account.  The reason why this is so is simply that
the state $|\vec p \sigma \rangle$,  which contains
a quasiparticle of momentum $\vec p$ and spin $\sigma$, is an
eigenstate of the current
operator $\hat{\vec j} = \sum_{i=1}^N\frac{\hat {\vec p}_i}{m_b}$ with
eigenvalue $\frac{\vec p}{m_b}$.
As a consequence, the current density associated with
the distribution $n_\sigma (\vec r, \vec p,t)$ is given by~\cite{footnote2}
\begin{eqnarray}\label{chargecurrents}
\vec j (\vec r , t) = \sum_{\vec p \sigma}
\frac{\vec p}{m_b} n_\sigma (\vec r,\vec p,t)~.
\end{eqnarray}

The difficulty in calculating the spin current arises 
from the fact that the state $|\vec p \sigma \rangle$ is {\it not} an eigenstate of
$\hat {\vec j}_{\uparrow}$ or  $\hat {\vec j}_{\downarrow}$: thus, we 
cannot automatically say that in this state $\vec j_\sigma = \frac
{\vec p}{m_b}$ and  $\vec j_{-\sigma} = 0$, even though these
expectation values would be consistent with the total value of
$\vec j$. All we can say, a priori, is that the expectation values
of  $\hat {\vec j}_{\uparrow}$  and $\hat {\vec j}_{\downarrow}$
in  the state $|\vec p \sigma \rangle$ must be proportional to
$\vec p$ and add up to $\frac{\vec p}{m_b}$.  Thus, we write
\begin{equation} \label{jtau}
\langle \vec p \sigma|\hat{\vec j}_{\tau}|\vec p \sigma\rangle
= \alpha_{\tau \sigma} \frac{\vec p}{m_b}~,
\end{equation}
where $\alpha_{\tau \sigma}$ is a 
 $2 \times 2$ matrix
whose columns add up to $1$, so that the total 
current is $\frac{\vec p}{m_b}$.  Notice that in a 
paramagnetic system $\alpha_{\uparrow \uparrow} 
= \alpha_{\downarrow \downarrow}$, and, 
therefore $\alpha_{\uparrow \downarrow} = \alpha_{\downarrow \uparrow}$:
for simplicity's sake,  we will focus on just this case from now on. 
The above Eq.~(\ref{jtau}) implies that the spin 
current carried by an up-spin quasiparticle of momentum $\vec p$ is
\begin{equation}   \label{js1}
\vec j_s (\vec p \uparrow) = (\alpha_{\uparrow \uparrow}
- \alpha_{\downarrow \uparrow})\frac{\vec p}{m_b}~ \equiv \frac{\vec p}{m_{s}}~,
\end{equation}
and, similarly, the spin-current carried by a down-spin quasiparticle
is
\begin{equation}   \label{js2}
\vec j_s (\vec p \downarrow) = (\alpha_{\uparrow \downarrow}
- \alpha_{\downarrow \downarrow})\frac{\vec p}{m_b}~\equiv  - \frac{\vec p}{m_{s}}~,
\end{equation}
since $\alpha_{\uparrow \uparrow} = \alpha_{\downarrow \downarrow}$.  
These equations define a {\it spin mass} $m_{s}$, which 
controls the spin current in much the  same way as $m_b$ controls 
the charge current~\cite{footnote3}.

Combining Eqs.~(\ref{js1}) and ~(\ref{js2}) we see that the correct expression 
for the spin current density carried 
by a nonequilibrium quasiparticle distribution $n_\sigma(\vec r,\vec p,t)$ is
\begin{eqnarray}   \label{js3}
\vec j_s (\vec r,  t ) = \sum_{\vec p} 
\frac{\vec p}{m_s} \left[n_\uparrow (\vec r , \vec p , t ) 
- n_\downarrow (\vec r , \vec p , t )\right]~.
\end{eqnarray}

\begin{figure} \label{fig1}
\includegraphics[width=7cm]{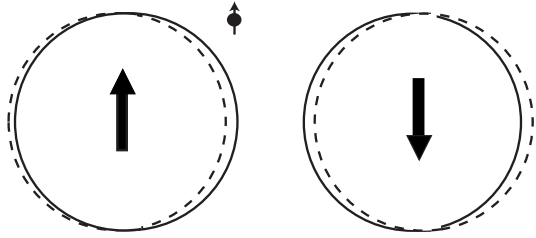}
\caption {Spin-momentum separation in a Fermi liquid:  
the momentum of an up-spin quasiparticle resides, 
in part, with down-spin particles.  The solid lines represent 
the Fermi surfaces and the dashed lines indicated the collective 
shifts in the momentum distribution of up- and down- spins.}
\vspace{1pt}
\end{figure}

It is clear that $m_s$ must be larger than $m_b$ since
$\alpha_{\uparrow \uparrow}$ and $\alpha_{\uparrow \downarrow}$
are positive numbers that add up to $1$,  implying that
$\alpha_{\uparrow \uparrow}-\alpha_{\downarrow \uparrow} <1$. The
positivity of  $\alpha_{\uparrow \uparrow}$ and $\alpha_{\uparrow
\downarrow}$ can be intuitively grasped by considering the
physical picture illustrated in Figure 1.  We start from an exact
eigenstate of the noninteracting system with full Fermi spheres of
up- and down-spins  and an additional single particle of momentum
$\vec p$ and spin $\uparrow$ out of the Fermi sphere. In this state $\vec
j_\uparrow =  \frac{\vec p}{m_b}$ and $\vec j_\downarrow = 0$.
The quasiparticle state is now obtained by slowly turning on the
electron-electron interaction.  The total momentum and spin do not
change in the process, but some momentum is transferred from the
up- to the down-spin component of the liquid:  one may say that
the up-spin quasiparticle drags along some down-spin electrons as
part of its ``screening cloud". As a result, the magnitude of
$\vec j_\uparrow$ is {\it smaller} than $\frac{\vec p}{m_b}$ by an
amount equal to $ {\vec j}_\downarrow$. The magnitude of the spin current
is {\it a fortiori} smaller than $\frac{\vec p}{m_b}$, which implies
$m_s>m_b$. Notice that the ``spin-momentum separation" described 
above is entirely due to correlations between 
electrons of opposite spin orientation.  Interactions between 
same-spin electrons do not contribute to this effect.

Having thus clarified the general concept of 
the spin mass we now proceed to (1) relate $m_s$ to the quasiparticle
effective mass and the Landau Fermi liquid parameters,  
(2) relate $m_s$ to the small wave vector and low frequency limit 
of the spin local field factor $G_-(q,\omega)$,  and (3) present
approximate microscopic calculations of $m_s$ in a paramagnetic 
electron liquid in three and two dimensions.

Let us start from the quasiparticle state $|\vec p \sigma \rangle$ and
apply to it the unitary transformation
$\hat U =\exp\left[i \sum_{i,\tau} \vec q_\tau \cdot \vec r_i
\frac{ 1 + \tau \hat \sigma_{z,i}}{2} \right]$, which 
boosts the momenta of the $\tau$-spin electrons  by  $\vec q_\tau$.
By applying $\hat U$ to the fundamental hamiltonian of the electron 
liquid  one can straightforwardly show
that the change in energy {\it of any} state, to first order in $\vec q_\sigma$,  is
\begin{equation}\label{DeltaE}
\Delta E =  \sum_\tau \vec j_\tau \cdot \vec q_\tau~.
\end{equation}
On the other hand, for the quasiparticle state under consideration,
we know that $\vec j_\tau = \alpha_{\tau \sigma} \frac{\vec p}{m_b}$.
Substituting this into Eq.~(\ref{DeltaE}) we get
\begin{equation}\label{DeltaE2}
\Delta E = \sum_\tau  \alpha_{\tau \sigma} \frac{ \vec p \cdot \vec q_\tau}{m_b}~.
\end{equation}
The energy change under this transformation can also be 
calculated with the help of the Landau theory of Fermi liquids.  
There are two contributions:  one from  the boost in the momentum of the quasiparticle, 
and the other from the collective  displacement of the Fermi 
surfaces by $\vec q_\tau$.  A standard calculation gives
\begin{equation}
\label{DeltaE3}
\Delta E = \sum_\tau \biggl \{ \frac{\vec p}{m^*} \delta_{\sigma \tau}
- \sum_{\vec p'} f_{\vec p \sigma, \vec p' \tau }\vec  
\nabla_{\vec p '} n_{0,\tau}({\vec p '})
\biggr \} \cdot {\vec q}_\tau~,
\end{equation}
where $n_{0,\tau}(\vec p) = \Theta (p_F-p)$ is the  momentum 
distribution in the ground state and $p_F$ is the Fermi momentum. 

Comparing Eqs.~(\ref{DeltaE2})
and~(\ref{DeltaE3}), we arrive at the identifications
\begin{eqnarray}
\alpha_{\uparrow \uparrow} &=& \frac{m_b}{m^*} 
\left[ 1+ \frac{F_1^{\uparrow \uparrow}}{2d} \right]~,\nonumber \\
\alpha_{\uparrow \downarrow} &=& \frac{m_b}{m^*} 
\frac{F_1^{\uparrow \downarrow}}{2d}~,
\end{eqnarray}
where $d$ is the number of spatial dimensions 
and $F_\ell^{\sigma \tau}  \equiv N^*(0) 
\int \frac{d \Omega}{4 \pi}f_{\vec p \sigma, 
\vec p' \tau } P_{\ell}(\cos \theta)$  is the angular average of 
the interaction function, weighted with the 
Legendre polynomial $P_\ell (\cos \theta)$ 
(or just $\cos \ell \theta$ in two dimensions) and multiplied 
by the density of states at the Fermi surface, $N^*(0)$.
Notice that the sum rule $\sum_{\tau}\alpha_{\tau \sigma}=1$
is satisfied by virtue of the well known Fermi liquid relation~\cite{NP}
\begin{equation}
\frac{m_b}{m^*} = \frac{1}{1 + F_{1}^s/d}~,
\end{equation}
where  $F_{\ell}^{s(a)}
= \frac{1}{2}\left[ F_{\ell}^{\uparrow \uparrow} 
+(-)F_{\ell}^{\uparrow \downarrow}\right]$ 
are the standard dimensionless Landau parameters  
defined, for example,  in Ref.~\cite{NP}.
The spin mass, on the other hand, is given by (see Eq.~(\ref{js1}))
\begin{equation} \label{spinmass1}
\frac{m_s}{m^*} = \frac{1}{1+F_{1}^a/d}~,
\end{equation}
showing that the relation of $m_s$ to $m^*$ is to 
the spin-channel what the relation of $m_b$ to $m^*$ is to the density channel.

It should be noted that the spin 
current density obtained from Eq.~(\ref{js3})  satisfies the 
continuity equation
\begin{eqnarray}    \label{continuity}
\frac{\partial }{\partial t} n_s (\vec r , t) 
+ \vec \nabla \cdot \vec j_s (\vec r , t) = 0~,
\end{eqnarray}
where $n_s ({\vec r}, t) = n_{\uparrow} ({\vec r}, t)
- n_{\downarrow} ({\vec r}, t)$, is the spin density.
Conversely, Eq.~(\ref{spinmass1}) could have been directly obtained 
from the requirements of charge and spin conservation.

The microscopic calculation of Landau parameters is notoriously
difficult.  Diagrammatic calculations of
$F_{1}^s$ and  $F_{1}^a$  in the three-dimensional 
electron liquid were done by Yasuhara and Ousaka~\cite{yasuhara}, and 
the calculated parameters, together with the resulting 
values of $m_s/m_b$ are listed in the upper half 
of Table~\ref{Table1} for various values of the
Wigner-Seitz radius $r_s$. In two dimensions the 
parameters $F_{1}^s$ and  $F_{1}^a$ were calculated by a 
variational Quantum Monte Carlo method in Ref.~\cite{kwon}.   
The parameters and the resulting values of $m_s/m_b$ are listed in 
the bottom half of Table~\ref{Table1}.  Notice that the spin mass 
enhancement in two dimensions is considerably higher than in three dimensions.

\begin{table}
\begin{tabular}{c c c c c c c}
\hline \hline  
$d$~~~   &   $r_s$  &    $ 1$   &   $2$   &
$3$ & $4$ &  $5$ \\  
\hline 
\\~$3$~~~~&$\frac{F_1^s}{d}~$ &    -0.0543~~    
&   -0.0647~~         &    -0.0713~~        &  -0.0773~~   & -0.0829~~  \\  
\\~$~$~~~~&$\frac{F_1^a}{d}~$ &    -0.0645~~    
&   -0.0825~~         &    -0.0915~~        &  -0.0956~~   & -0.0965~~  \\ 
\\~$~$~~~~&$\frac{m_s}{m_b}~$ &    1.011~~    
&   1.019~~         &    1.022~~        &  1.020~~   & 1.015~~  \\ \\  
\hline \hline
\\~$2$~~~~&$\frac{F_1^s}{d}~$ &    -0.071~~    
&   -0.050~~ &    -0.015~~ & ~--~ & 0.061~ \\  
\\~$~$~~~~&$\frac{F_1^a}{d}~$ &    -0.096~~    
&  -0.120~~         &    -0.130~~        & ~--~   & -0.136~~  \\ 
\\~$~$~~~~&$\frac{m_s}{m_b}~$ &1.028 &1.080 &1.132&~--~ &1.228\\ \\
\hline \hline
\end{tabular}
\caption{Landau parameters from Refs.~\cite{yasuhara} 
and~\cite{kwon} and spin mass enhancement 
$\frac{m_s}{m_b} = \frac{1+F_1^s/d}{1+F_1^a/d}$ in the d-dimensional electron liquid.} 
\label{Table1}  
\end{table}

In view of the uncertainty in the calculation of  
the Landau parameters it seems worthwhile  to attempt 
another kind of calculation, which does not rely  on diagrammatic expansions.  
We first establish the
connection between the spin mass and the dynamical {\it local field factor}  
in the spin channel.  We recall that the dynamical spin susceptibility of
an electron liquid  is usually represented in the form
\begin{equation} \label{chispin}
\chi_s (q, \omega) = \frac{\chi_0(q, \omega)}{1 +v_qG_-(q,\omega) \chi_0(q,\omega)}~,
\end{equation}
where $\chi_0(q,\omega)$ is the noninteracting spin susceptibility 
(i.e., the Lindhard function), $v_q$ is the Fourier transform of 
the Coulomb interaction ($=4 \pi e^2/q^2$ in three dimensions 
and $2 \pi e^2/q$ in two dimensions) and
$G_-(q,\omega)$ is the dynamical  local field factor in the spin channel.
In the limit $q \to 0$ and  small, but finite 
frequency ($\omega \ll \epsilon_F/\hbar$ 
where $\epsilon_F$ is the Fermi energy),  Eq.~(\ref{chispin}) reduces to
\begin{eqnarray} \label{chimacro1}
\chi_s(q,\omega) \stackrel{q \to 0} {\to} 
\frac{n q^2}{m_b  \left[1 +  \lim_{q \to 0}
\frac{n q^2v_q G_-(q,\omega)}{m_b \omega^2} \right] \omega^2}~.\nonumber \\
\end{eqnarray}
 On the other hand,  the small-$q$/finite-$\omega$ limit 
of $\chi_s(q,\omega)$ can also be
calculated by solving the kinetic equation~\cite{NP}
in the presence of  slowly varying external fields $V_{\sigma}(q,\omega)$.
In this region collisions are irrelevant, and one gets the spin response
\begin{eqnarray}
\delta n_s (\vec q, \omega) \simeq
 \frac{n q^2}{m_s \omega^2}~V_s (\vec q, \omega)~,
\end{eqnarray}
where $\delta n_s (\vec q, \omega) =  \delta n_\uparrow (\vec q, \omega)
- \delta n_\downarrow (\vec q, \omega) $ 
and $V_s (\vec q, \omega) = [ V_\uparrow (\vec q, \omega)
- V_\downarrow (\vec q, \omega) ]/2 $.
Therefore,
\begin{equation} \label{chimacro2}
\chi_s(q,\omega) \stackrel{q \to 0}{\to} \frac{n q^2}{m_s \omega^2}~.
\end{equation}
Comparing the above equation with Eq.~(\ref{chimacro1}) leads to the identification
\begin{equation}  \label{spinmass2}
\frac{m_s}{m_b} = 1 +  \lim_{\omega \to 0} 
\lim_{q \to 0}\frac{n q^2 v_qG_-(q,\omega)}{m_b \omega^2}~.
\end{equation}
The order of the limits is, of course, essential.  
When $\omega$ tends to zero first, $G_-(q,\omega)$ 
vanishes as $q^{d-1}$ for $q \to 0$, so as to yield a finite 
enhancement of the uniform static spin susceptibility.  
In Eq.~(\ref{spinmass2}), however, $q$ tends to zero first, 
and we see that $v_qG_-(q,\omega)$ must 
go as $\frac{\omega^2}{q^2}$ in order to give a finite value of the spin mass. 

The above analysis, combined with the Kramers-Kr\"onig dispersion 
relation, leads to the following relation 
between the real and the imaginary part of $G_-(q,\omega)$:
\begin{equation}\label{KK}
\lim_{q \to 0}\Re e G_-(q,\omega) 
=  \lim_{q \to 0}\frac{2 }{\pi}{\cal P}\int_0^\infty d \omega' 
~\frac{\omega^2 \Im m G_-(q,\omega')}{\omega'(\omega'^2 - \omega^2)}~,
\end{equation}
where ${\cal P}$ denotes the principal-part.    
In the $\omega \to 0$ limit, comparison with Eq.~(\ref{spinmass2}) yields
\begin{equation} \label{spinmassmicro}
\frac{m_s}{m_b} = 1 + \frac{n}{m_b}\lim_{q \to 0}\frac{2}{\pi}
\int_0^\infty d \omega ~\frac{q^2 v_q~\Im m G_-(q,\omega)}{\omega^3}~.
\end{equation}

The quantity $\lim_{q \to 0} q^2 v_q \Im m G_-(q,\omega)$ was 
written as a convolution of the response 
functions $\chi_{\sigma \sigma'}(k,\omega)$
in the mode-decoupling theory \cite{Qian02,Hasegawa69,Nifosi98,QCV} 
in 3-d in Ref. \cite{QCV}, (in which it was denoted as $A(\omega)$,
and note that the prefactor $\frac{4}{3V}$ 
becomes $\frac{4}{dL^d}$ in $d$ dimensions, 
with $L$ the linear size of the system). 
The results of our
calculations of the spin mass from Eq. (\ref{spinmassmicro}), with
the response functions $\chi_{\sigma \sigma'}(k,\omega)$ evaluated in the
generalized random phase approximation (GRPA), are listed in 
Table~\ref{Table2}. The static local field factors in GRPA
are taken from Ref.~\cite{iwamoto1} in 3-d,
and from Refs.~\cite{iwamoto2}  
and~\cite{davoudi} in 2-d, respectively.
Although there are considerable
differences between the numbers obtained in different approximations,
we see that the  values of the spin mass obtained by
this method are  consistently larger  than the ones listed in Table~\ref{Table1}.

\begin{table}
\begin{tabular}{c c c c c c c}
\hline \hline 
$d$~~~   &   $r_s$  &    $ 1$   &   $2$   &
$3$   &   $4$ &  $5$\\
\hline
\\~$3$~~~~&$\frac{m_s}{m_b}~~$ &  1.02~~&1.06~~&    1.11~~         
&   1.17~~   & 1.23~~ \\ 
\\~$~~$~~~~&$\frac{m_s}{m_b}~~$ &  1.01~~&1.03~~&    1.03~~         
&   1.04~~   & 1.04~~ \\ \\
\hline \hline
\\~$2$~~~~&$\frac{m_s}{m_b}~~$ &  1.15~~ &1.46~~ &1.83~~ 
&2.21 ~~ &2.59~~ \\ 
\\~$~$~~~~&$\frac{m_s}{m_b}~~$ &  1.18~~ &1.77~~ &2.78~~ 
&4.11 ~~ &5.36~~
 \\  \\
\hline \hline
\end{tabular}
\caption{Spin mass enhancement calculated from Eq. (\ref{spinmassmicro}). 
The first, third, and fourth lines are 
calculated with $g(\omega)=1$ \cite{QCV,footnote5}, 
while the second line includes an empirical 
correction  for the third moment sum rule in 3-d \cite{QCV,Goodman}. The 
local field factors are taken from Ref.~\cite{iwamoto1} in 3-d,   
and from Refs.~\cite{iwamoto2} (third line) and~\cite{davoudi} (fourth line) in 2-d.}  
\label{Table2}
\end{table}

In summary, we have shown that the calculation
of the spin-current in an electronic system is a delicate task:
it is not sufficient to include interactions in 
the transport equation for the quasiparticle distribution 
function: one must also use the correct spin mass to calculate 
the spin-current from the distribution function. 
We have found that the difference between the spin mass 
and the bare band mass  is much larger in two dimensional systems than in three-dimensional ones. 
Although the spin masses calculated in various schemes in 2-d
are quite different from each other and might be overestimated in some
cases due to the limitations of the approximations employed, there is no doubt that they all indicate a significant many-body effect which is definitely large enough to be observable in the exciting practice of 2-d
spintronics.  We hope that these results will stimulate more accurate 
calculations of the spin mass by quantum Monte Carlo methods.

We gratefully acknowledge support by NSF grants DMR-0074959 and DMR-0313681  
and by DOE grant DE-FG02-01ER45897.


\begin{thebibliography}{100}
\bibitem{footnote2}
A nice discussion of the physical
origin of the difference between the quasiparticle current and its group
velocity $\frac{\vec p}{m^*}$ can be found in Ref.~\cite{NP}.
\bibitem{footnote3}In a spin-polarized system the eigenvalues of 
$\alpha_{\sigma \sigma'}$ still define two masses $m_b$ and $m_s$:  
while $m_b$ is still associated with the charge channel, $m_s$ is 
now associated with a linear combination of the charge and the spin channel.
\bibitem{footnote5} In three dimensions this sum rule 
tells us that $\lim_{\omega \to \infty}\lim_{q \to 0}q^2v_q\Re e G_-(q,\omega) 
= \frac{8 \pi e^2}{3}[g_{\uparrow \downarrow}(0)-1]$,~\cite{Goodman}   
where $g_{\uparrow \downarrow}(0)$ is the pair correlation 
function for opposite spin electrons at zero separation,  
while in two dimensions we find the same quantity to diverge logarithmically.
\end{thebibliography}
\end{document}